\documentclass[aps,preprint,prd,showpacs,nofootinbib]{revtex4}

\usepackage{afterpage}
\usepackage[utf8]{inputenc}

\usepackage[T1]{fontenc} 
\usepackage{latexsym}
\usepackage{amsmath}
\usepackage{graphicx}
\usepackage{hyperref}
\usepackage{subfigure}
\usepackage{dcolumn}
\usepackage{bm}
\usepackage{amssymb}
\usepackage{latexsym}
\usepackage{datetime}
\usepackage{scrtime}
\usepackage{color}
\usepackage[dvipsnames, svgnames, x11names]{xcolor}
\usepackage{float}

\hypersetup{colorlinks=true,
	breaklinks=true,
	pdfstartview=Fit,
	linkcolor=blue,
	citecolor=blue,
	urlcolor=blue}

\def\be{\begin{equation}}
\def\ee{\end{equation}}
\def\ba{\begin{eqnarray}}
\def\ea{\end{eqnarray}}

\def\lf{\left}
\def\rt{\right}
\def\bal{\begin{aligned}}
\def\eal{\end{aligned}}
\def\d{\mathrm{d}}

\begin{document}
	

\title{Parity-violating primordial gravitational waves from null energy condition violation}

\author{Zi-Wei Jiang$^{1}$}
\email[]{jiangzw@gs.zzu.edu.cn}
\author{Yong Cai$^{1}$}
\email[]{caiyong@zzu.edu.cn}
\author{Fei Wang$^{1}$}
\email[]{feiwang@zzu.edu.cn}
\author{Yun-Song Piao$^{2,3,4,5}$}
\email[]{yspiao@ucas.ac.cn}

\affiliation{$^1$ \small{Institute for Astrophysics, School of Physics, Zhengzhou University, Zhengzhou 450001, China}}
\affiliation{$^2$ \small{School of Physical Sciences, University of Chinese Academy of Sciences, Beijing 100049, China}}
\affiliation{$^3$ \small{International Centre for Theoretical Physics Asia-Pacific, University of Chinese Academy of Sciences, 100190 Beijing, China}}
\affiliation{$^4$ \small{School of Fundamental Physics and Mathematical Sciences, Hangzhou Institute for Advanced Study, UCAS, Hangzhou 310024, China}}
\affiliation{$^5$ \small{Institute of Theoretical Physics, Chinese Academy of Sciences, P.O. Box 2735, Beijing 100190, China}}

\begin{abstract}

We investigate the parity-violating effects in primordial gravitational waves (GWs) due to null energy condition (NEC) violation in two very early universe scenarios: bounce-inflation and intermediate NEC violation during inflation. In both scenarios, we numerically solve the power spectra of parity-violating primordial GWs generated by coupling the background field and the spectator field with the Nieh-Yan term, respectively.
We find that the background field can significantly enhance parity-violating effects at scales corresponding to the maximum of the GW power spectra. In contrast, the parity-violating effects produced by the spectator show significantly weaker observability even if the coupling constant is large.
Therefore, in NEC-violating scenarios, the significant observable parity-violating effects in primordial GWs primarily arise from the physics directly related to NEC violation. This result highlights the potential of primordial GWs as crucial tools for exploring NEC-violating and parity-violating physics.

\end{abstract}

\maketitle
\tableofcontents

\section{Introduction}
\label{sec:intro}

LIGO's detection of gravitational waves (GWs) \cite{LIGOScientific:2016aoc} from the merger of binary black holes validated a crucial prediction of general relativity made a century ago. In 2023, several pulsar timing array (PTA) collaborations, including NANOGrav \cite{NANOGrav:2023hvm,NANOGrav:2023gor}, EPTA \cite{EPTA:2023fyk}, PPTA \cite{Reardon:2023gzh}, and CPTA \cite{Xu:2023wog}, presented compelling evidence of a signal consistent with stochastic GW background at nanohertz frequencies. These findings have significantly advanced the study of GWs and the exploration of the universe.

Inflation \cite{Guth:1980zm,Linde:1981mu,Albrecht:1982wi,Starobinsky:1980te} is the leading paradigm of the early universe. The primordial curvature perturbations predicted by slow-roll inflation are consistent with observations of the temperature anisotropy of cosmic microwave background (CMB) \cite{Planck:2018vyg,Planck:2018jri}. At the same time, slow-roll inflation predicts nearly scale-invariant primordial GWs, with current observations at the scale of the CMB placing an upper bound on the tensor-to-scalar ratio, i.e., $r_{0.002}<0.035$ \cite{BICEP:2021xfz}. Primordial GWs cover a broad frequency range (about $10^{-18} - 10^{10}$Hz) and may carry rich information about the early universe and gravitational interactions at high-energy scales.

Parity violation is well-established in weak interaction \cite{Lee:1956qn}, but it is absent in strong and electromagnetic interactions. Although GWs are parity-conserving in general relativity, they exhibit asymmetry between their left and right-handed modes in parity-violating theories of gravity \cite{Lue:1998mq,Alexander:2004wk,Satoh:2007gn}, see e.g., studies in gravitational Chern-Simons gravity \cite{Jackiw:2003pm,Alexander:2009tp,Cai:2016ihp,Bartolo:2017szm,Yoshida:2017cjl,Kawai:2017kqt,Bartolo:2018elp,Nishizawa:2018srh,Nojiri:2019nar,Fujita:2020iyx,Nojiri:2020pqr,Chu:2020iil,Fu:2020tlw,Odintsov:2022hxu,Peng:2022ttg,Cai:2022lec,Zhu:2023lhv,Fronimos:2023tim,Kanno:2023kdi} and Nieh-Yan modified teleparallel gravity \cite{Li:2020xjt,Li:2021wij,Cai:2021uup,Wu:2021ndf,Li:2023fto,Fu:2023aab}, see also \cite{Li:2022mti,Li:2022vtn,Bastero-Gil:2022fme,Qiao:2022mln,Rao:2023doc,Zhang:2024vfw,Hu:2024hzo,Akama:2024bav} for recent studies.
Whether there is a parity violation in gravitational interactions remains to be experimentally verified, see e.g. investigations utilizing CMB polarization data \cite{Minami:2020odp,Diego-Palazuelos:2022dsq,Eskilt:2022cff,Fujita:2022qlk}, as well as future observations from ground- and space-based interferometers \cite{Smith:2016jqs,Yagi:2017zhb,Domcke:2019zls,Martinovic:2021hzy,Jiang:2022uxp}.

Primordial GW could serve as a crucial tool for testing the parity-violating effect of gravitational interactions at high-energy scales. However, for typical slow-roll inflation models, the amplitude of the predicted nearly scale-invariant primordial GW power spectrum is constrained by CMB data \cite{BICEP:2021xfz}. Additionally, in models where the inflaton is coupled to a parity-violating term, the effect of parity violation is suppressed by the slow-roll condition.
For these reasons, even if primordial GWs exhibit parity violation, this effect is unpromising to be detected by ground- or space-based interferometers or the PTA.

However, the evolution of the very early universe may be much more complex than that of slow-roll inflation. A violation of the null energy condition (NEC) could play a crucial role in the origin and evolution of the universe at its very early stages. Bounce-inflation \cite{Piao:2003zm,Piao:2005ag,Piao:2004me,Liu:2013kea,Qiu:2015nha,Li:2016awk,Cai:2016thi,Cai:2017dyi,Cai:2017pga} (or Genesis-inflation \cite{Liu:2014tda,Pirtskhalava:2014esa,Kobayashi:2015gga,Cai:2017tku}), which violates the NEC during certain stage, can address the cosmological singularity problem \cite{Borde:1993xh,Borde:2001nh} (see also \cite{Cai:2019hge}) and explain the large-scale anomaly in the CMB \cite{Piao:2003zm,Piao:2005ag,Liu:2013kea,Cai:2015nya}. It has been found in \cite{Cai:2020qpu,Cai:2022nqv} that an intermediate NEC violation during inflation can significantly enhance the amplitude of the primordial GW power spectrum, potentially explaining the signal of stochastic GW background observed at PTA scales \cite{Ye:2023tpz} (see also \cite{Jiang:2023gfe,Zhu:2023lbf,Pan:2024ydt,Chen:2024mwg,Chen:2024jca,Choudhury:2023hfm,Choudhury:2023kam,Choudhury:2024one,Choudhury:2023fjs}).

In these scenarios, if the background field is coupled to a parity-violating term, the NEC violation naturally amplifies the parity-violating effects in primordial GWs \cite{Cai:2022lec,Zhu:2023lhv}. This is because NEC violation naturally disrupts the slow-roll condition. Additionally, NEC violation during inflation can also significantly enhance the primordial curvature perturbation on small scales, potentially leading to the formation of primordial black holes and scalar-induced GWs at certain scales \cite{Cai:2023uhc}.
A combination of these observational effects could serve as a valuable probe for exploring the NEC violation in the very early universe.

In this paper, we examine the effect of NEC violation on enhancing the parity-violating effects in primordial GWs within two scenarios: bounce-inflation and inflation with intermediate NEC violation. In both scenarios, we investigate the coupling of the background field (see, e.g., \cite{Wang:2014abh, Cai:2022lec}) and the spectator field with the parity-violating term to discern the differences in how NEC violation impacts the enhancement of parity-violating effects in primordial GWs. To get rid of the ghost problem, we set the parity-violating term as the Nieh-Yan term, see e.g.,  \cite{Li:2020xjt,Li:2022mti}.

The rest of this paper is organized as follows. In Sec. \ref{section2}, we compare the parity-violating effects on primordial GWs when the background field and the spectator field are coupled with the Nieh-Yan term, respectively, within the bounce-inflation scenario. In Sec. \ref{3}, parallel to Sec. \ref{section2}, we conduct a similar investigation in the scenario of inflation with intermediate NEC violation. In Sec. \ref{Sec:Discuss}, we discuss several key points closely related to this work. Sec. \ref{conclusion} provides our conclusions.
In all the numerical calculations throughout the paper, we set ${M_{\text{P}}}^2 \equiv (8\pi G)^{-1} = 1$.

\section{Parity-violating primordial GWs in bounce-inflation scenario} \label{section2}

In this section, we will investigate the presence of parity-violating primordial GWs in bounce-inflation models, considering scenarios where the background field and spectator field coupled with the parity-violating term (i.e., the Nieh-Yan term).
We will obtain the power spectrum of primordial GWs through numerical calculations. Our focus will be on discerning the differences in how the background field and the spectator field impact parity-violating effects.

\subsection{Setup}

We begin with the following effective action
\be
S=\int \d^4x\sqrt{-g}\lf[{{M_{\text{P}}}^2 \over 2}R -{{M_{\text{P}}}^2 \over 2}X-V\lf(\phi\rt)+\tilde{P}\lf(\phi,X\rt) +\mathcal{L}_{\psi}
+\mathcal{L}_{NY}\rt]\,, \label{binfaction}
\ee
where $R$ is the Ricci scalar, $\phi$ is the dimensionless scalar field that determines the background evolution, $X = \partial_{\mu}\phi\partial^{\mu}\phi$, and $\tilde{P}(\phi,X)$ is a function of $\phi$ and $X$.
We assume that the background evolution is dominated by $\phi$.
We set $\mathcal{L}_{\psi}=0$ when considering the coupling of the background field with the Nieh-Yan term. When considering the coupling of the spectator field with the Nieh-Yan term, we introduce the axion as the spectator field $\psi$, described by $\mathcal{L}_{\psi}=-\nabla_{\mu}\psi\nabla^{\mu}\psi/2-U(\psi)$.
The term $\mathcal{L}_{NY}$ represents the coupling of $\phi$ or $\psi$ with the Nieh-Yan term, i.e.,
\be
\mathcal{L}_{NY}={\xi F(\phi,\psi)\over 4}\mathcal{T}_{A\mu\nu}\tilde{\mathcal{T}}^{A\mu\nu}\,,\label{NYphi}
\ee
where $\tilde{\mathcal{T}}^{A\mu\nu}= \epsilon^{\mu\nu\rho\sigma}{\mathcal{T}^A}_{\rho\sigma}/2$ is the Hodge dual of torsion two form, $\epsilon^{\mu\nu\rho\sigma}$ is the four-dimensional Levi-Civita tensor and $\xi$ is the coupling constant, see \cite{Li:2020xjt,Li:2021wij} for details (see also e.g. \cite{Gialamas:2022xtt,Gialamas:2023emn}).
We will set the coupling function as $F=\phi$ and $F=\psi$ in the two different coupling scenarios.
It is worth noting that the background evolution remains unaffected by $\mathcal{L}_{NY}$ \cite{Li:2020xjt, Li:2021wij}.

In this paper, we will not address the primordial scalar perturbation. In fact, to ensure its stability throughout the entire evolutionary history of the universe, we need to incorporate the effective field theory operator $L_{\delta g^{00}R^{(3)}}$ into the action \cite{Cai:2017dyi}. Fortunately, this operator does not affect the background evolution or the primordial GWs.

The background evolution of bounce-inflation can be constructed with (see \cite{Cai:2017pga})
\be
\tilde{P}(\phi,X)
=\alpha(\phi){M_{\text{P}}^2 \over 2}X
+\beta(\phi){X^2 \over 4},
\ee
\ba\bal
V(\phi)=-{V_0 \over 2}e^{\sqrt{2 \over q}\phi}\lf[1-\tanh\lf({\phi \over \lambda_2}\rt)\rt]
+{\Lambda \over 2}\lf[1-\lf({\phi \over \lambda_3}\rt)^2\rt]^2\lf[1+\tanh\lf({\phi \over \lambda_2}\rt)\rt].
\eal\ea
where
\be
\alpha(\phi)=\alpha_0 \lf[1+\lf({\phi \over \lambda_1}\rt)^2\rt]^{-2} \,, ~~~
\beta(\phi)= \alpha(\phi)\beta_0/\alpha_0 \,,
\ee
$\lambda_{1,2,3}$, $q$, $\alpha_0$ and $\beta_0$ are positive dimensionless constants.
Based on this construction, the universe originates from an ekpyrotic contraction \cite{Khoury:2001wf} and enters inflation after a nonsingular bounce.

With the line element $\d s^2=-\d t^2+a^2(t)\d{\vec{x}}^2$, the Friedmann equations can be given as
\be
3H^2M_{\text{P}}^2=\lf[1-\alpha\lf(\phi\rt)\rt]{M_{\text{P}}^2 \over 2}{\dot{\phi}}^2
+{3 \over 4}\beta\lf(\phi\rt){\dot{\phi}}^4+V\lf(\phi\rt)+\rho_{\psi}\,,
 \ee
\be
\dot{H}M_{\text{P}}^2=-\lf[1-\alpha\lf(\phi\rt)\rt]{M_{\text{P}}^2 \over 2}\dot{\phi}^2-{1 \over 2}\beta\lf(\phi\rt)\dot{\phi}^4 - {1\over2}(\rho_\psi+p_\psi) \,,
\ee
where a dot denotes ``${\d / \d t}$'', $H$ is the Hubble parameter, $\rho_\psi$ and $p_\psi$ represent the energy density and pressure of the spectator field $\psi$, respectively.
For our purpose, we require that $\rho_\psi$ and $p_\psi$ are subdominant in the Friedmann equations.

\begin{figure}[htbp]
  \centering
  \hspace{0mm}
  \subfigure[$H$]{\includegraphics[width=0.46\textwidth]{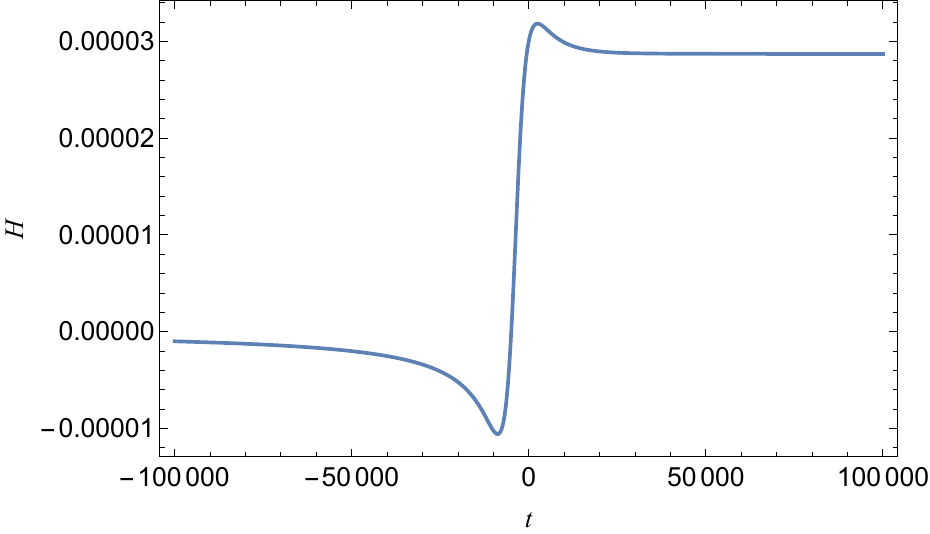}}
  \hspace{0mm}
  \subfigure[$\epsilon\equiv -\dot{H}/H^2$]{\includegraphics[width=0.43\textwidth]{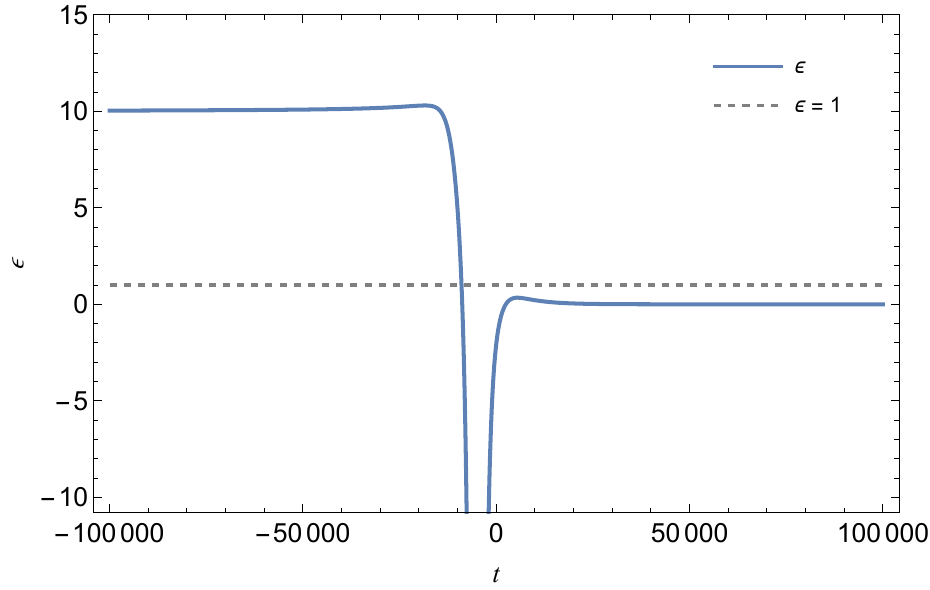}}
   \hspace{0mm}
  \subfigure[$\phi$]{\includegraphics[width=0.45\textwidth]{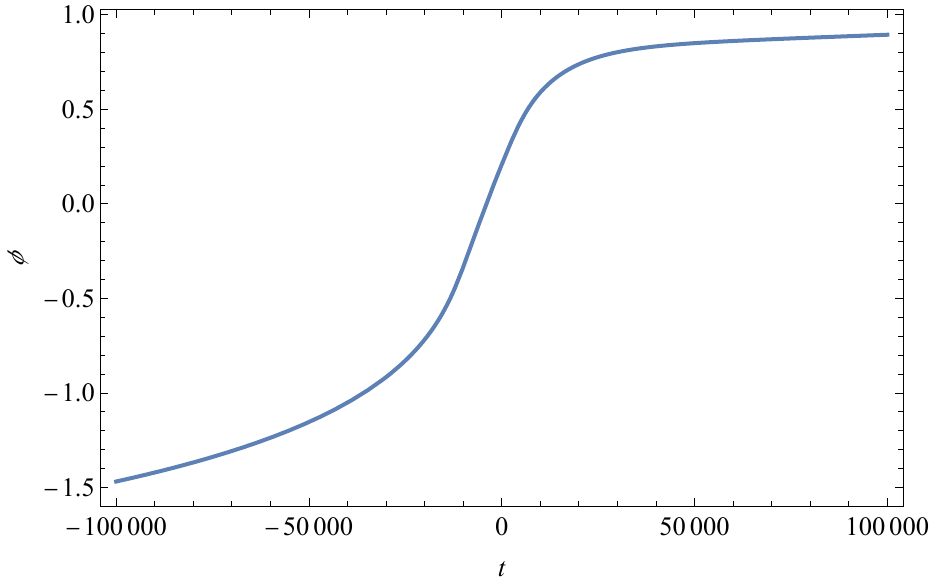}}
  \hspace{0mm}
  \subfigure[$\dot{\phi}$ and $\phi'$]{\includegraphics[width=0.45\textwidth]{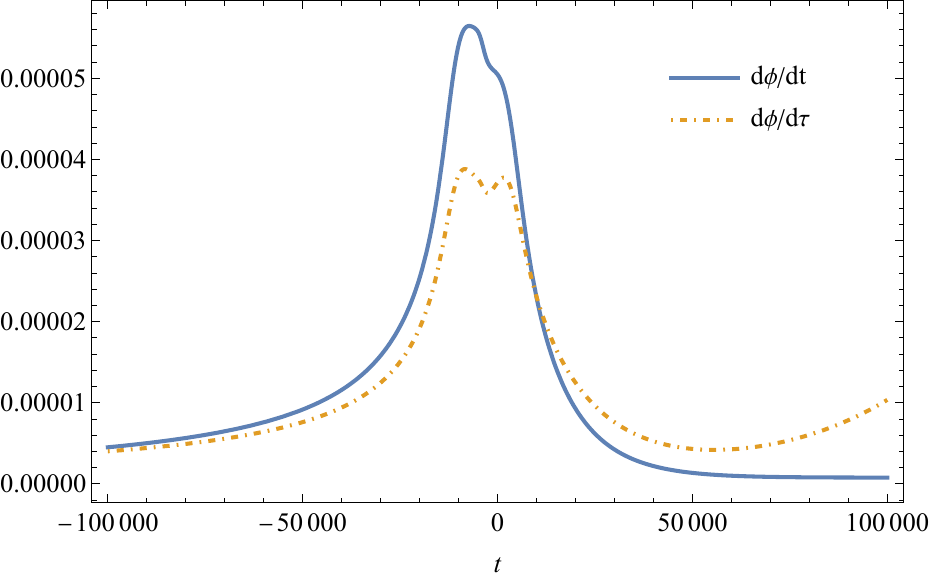}}
  \caption{The background evolution in bounce-inflation model, where we have set $\alpha_0=20$, $\beta_0=4.5\times 10^9$, $\lambda_1=0.224$, $V_0=5\times10^{-9}$, $q=0.1$, $\lambda_2=0.067$, $\lambda_3=12$, $\Lambda=2.5\times10^{-9}$.}
  \label{figbinfbg}
\end{figure}

Following \cite{Cai:2017pga}, we implement the background evolution of bounce-inflation by setting
$\alpha_0=20$, $\beta_0=4.5\times 10^9$, $\lambda_1=0.224$, $V_0=5\times10^{-9}$, $q=0.1$, $\lambda_2=0.067$, $\lambda_3=12$, $\Lambda=2.5\times10^{-9}$. After numerically solving the Friedmann equations, we illustrate the background evolution of bounce-inflation in Fig. \ref{figbinfbg} by plotting $H$, $\epsilon\equiv -\dot{H}/H^2$, $\phi$ and $\dot{\phi}$ with respect $t$. As shown in Figs. \ref{figbinfbg} (a) and (b), the universe originates from an ekpyrotic contraction ($H<0$ and $\epsilon\gg1$) and transitions into slow-roll inflation ($H\approx \text{const.}>0$ and $|\epsilon|\ll1$) after going through a NEC-violating nonsingular bounce ($\dot{H}>0$). The NEC violation significantly amplifies the Hubble parameter $H$ and the rate of change of the background field $\phi$ (i.e., $\dot{\phi}$). Consequently, the effect of parity violation can be naturally enhanced by the NEC violation, as we will demonstrate in the subsequent sections, assuming $F=\phi$.


In momentum space, the equation of motion for the primordial tensor perturbation (i.e., the primordial GW) mode ${h}^{(s)}_k$ is given by (see e.g. \cite{Li:2020xjt} for the derivation)
\be
\ddot{h}^{(s)}_k+3H\dot{h}^{(s)}_k+{k \over a}\left[{k \over a}+\lambda^{(s)}\xi\dot{F}\right]h^{(s)}_k=0 \,, \label{hk}
\ee
where $s=R$ and $L$ represent the left- and right-handed polarization modes, respectively, with $\lambda^{(L)}=-1$ and $\lambda^{(R)}=+1$. For convenience, we also define $s=N$ for the situation without parity violation, i.e., $\lambda^{(N)}=0$.
After defining $u_k^{(s)}=ah_k^{(s)}$, Eq. (\ref{hk}) can be rewritten as
\be
{u_k^{(s)}}''+\lf[{c_T^{(s)}}^2 k^2-{a'' \over a}\rt]{u_k^{(s)}}=0 \,, \label{uk}
\ee
where a prime denotes $\d/\d\tau$, $\d\tau=\d t/a$, ${c_T^{(s)}}^2\equiv 1+\mu^{(s)}$ and $\mu^{(s)}=\lambda^{(s)}\xi F'/k$ represents the parity violation.

Initially, the perturbation modes are deep inside their horizon, i.e., ${w^{(s)}}^2\gg a''/a$. The initial condition can be set as $u_k^{(s)}\simeq e^{-iw^{(s)}\tau}/\sqrt{2w^{(s)}}$.
For convenience, we assume that the parity violation effect is negligible at the initial time. Therefore, we would have $w^{(s)}= k$ in the limit $\tau\rightarrow-\infty$.



The power spectrum of the primordial GWs is defined as
\be P_T^{(s)}\equiv\frac{k^3}{2\pi^2}\lf|h_k^{(s)}\rt|^2 \,,\label{eq:power240601}
\ee
which is evaluated at the end of inflation. The total power spectrum of both left- and right-handed GW modes is given by $P_T=P_T^{(L)}+P_T^{(R)}$.
Additionally, we can introduce the chirality parameter as
\be
\Delta\chi \equiv \frac{P_T^{(L)}-P_T^{(R)}}{P_T^{(L)}+P_T^{(R)}}\,,\label{eq:Chi}
\ee
which can be used to assess the degree of parity violation in primordial GWs.

\subsection{Coupling the background field with the parity-violating term}\label{2.1}

In this subsection, we investigate the parity-violating primordial GWs in a single-field model of bounce-inflation and set ${\cal L}_\psi=0$ in Eq. (\ref{binfaction}). Therefore, we have $\rho_\psi=p_\psi=0$. The background evolution is displayed in Fig. \ref{figbinfbg}, see also \cite{Cai:2017pga}. We can see that $\dot{\phi}$ reaches its maximum around the NEC-violating phase. The background field is non-minimally coupled with the Nieh-Yan term, i.e., $F=\phi$ in Eq. (\ref{NYphi}).
Consequently, the background evolution determines not only the GW power spectrum $P_T$ but also the parity violation effect $\Delta\chi$.

Since the primordial GW power spectrum approximately satisfies $P_T\propto H^2$, a significant growth of $H$ during NEC violation will lead to a significant enhancement of $P_T$. The parity-violating effect $\Delta \chi$ of primordial GWs depends on $\dot{\phi}$ and the coupling constant $\xi$.
In the bounce-inflation model, for the coupling of the background field with the parity-violating term, the power spectrum of primordial GWs and the effect of parity violation were calculated in \cite{Li:2016awk} and \cite{Wang:2014abh}, respectively.
In this paper, our main focus is on the differences in the parity violation effects on primordial GWs when the background field $\phi$ and the spectator field $\psi$ are coupled to the parity-violating term, respectively.

\begin{figure}[htbp]
  \centering
  \hspace{0mm}
  \subfigure[]{\includegraphics[width=0.32\textwidth]{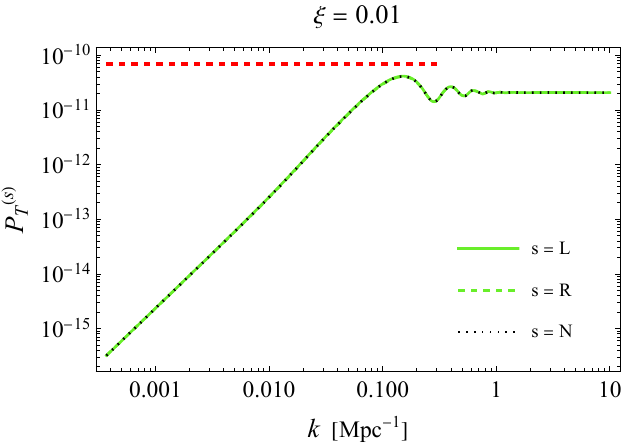}}
  \hspace{0mm}
  \subfigure[]{\includegraphics[width=0.32\textwidth]{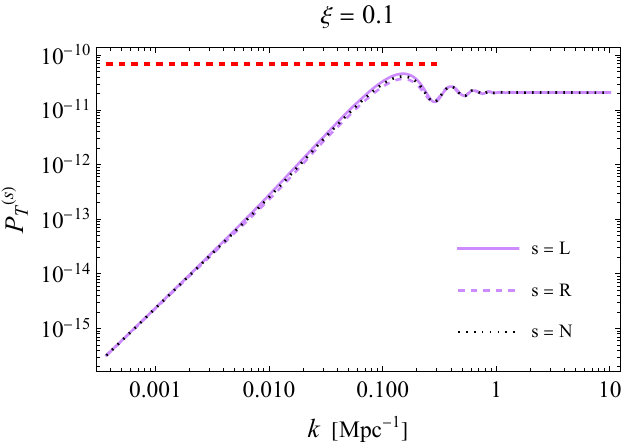}}
  \hspace{0mm}
  \subfigure[]{\includegraphics[width=0.32\textwidth]{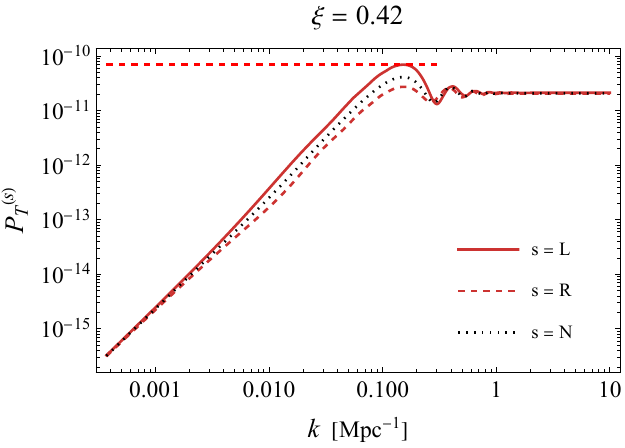}}
  \caption{The primordial GW power spectra $P_T^{(s)}$ for the left-handed ($s=L$, solid curves) and right-handed ($s=R$, dashed curves) modes in the bounce-inflation model for different values of $\xi$ when the background field $\phi$ is coupled with the Nieh-Yan term. For comparison, we also present the GW power spectra without parity violation ($s=N$, black dotted curves), corresponding to the case when $F \equiv 0$. The red horizontal dashed line denotes the constraint on the tensor-to-scalar ratio, i.e., $r_{0.002}<0.035$.}
  \label{figbinfoneAPT}
\end{figure}

\begin{figure}[htbp]
  \centering
  \hspace{0mm}
  \subfigure[]{\includegraphics[width=0.43\textwidth]{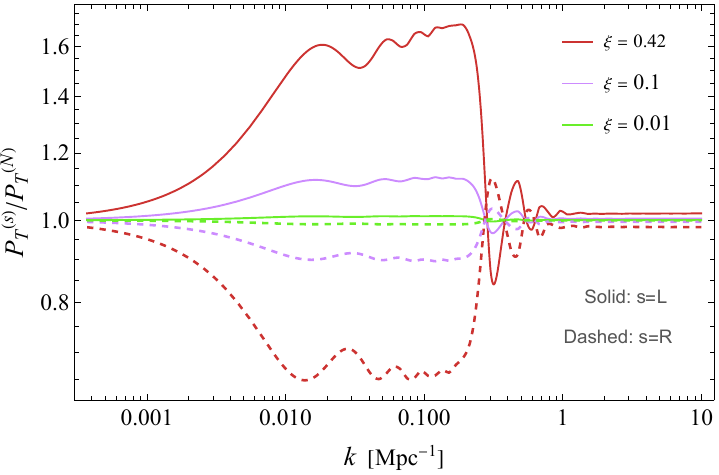}}
  \hspace{0mm}
  \subfigure[]{\includegraphics[width=0.435\textwidth]{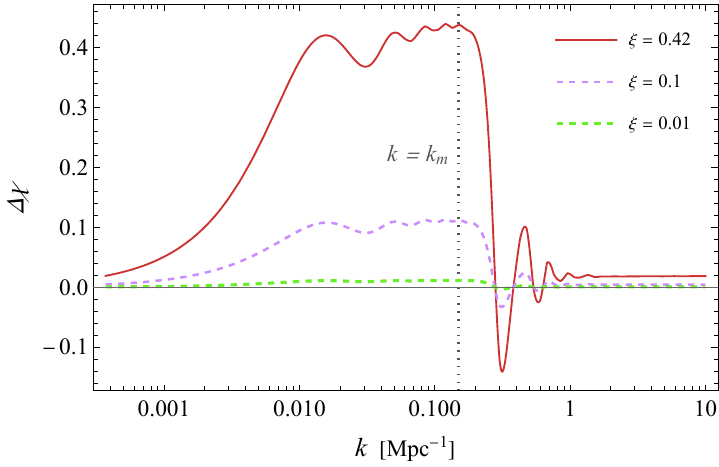}}
  \caption{The ratio of the primordial GW power spectra with and without parity violation ($P_T^{(s)}/P_T^{(N)}$) for different values of $\xi$ in the bounce-inflation model when the background field $\phi$ is coupled with the Nieh-Yan term, where $s=L$ and $R$ for the solid and dashed curves, respectively. The chiral parameter $\Delta \chi$ is shown in the right panel, where $k=k_m$ approximately corresponds to the wavenumber at which the power spectrum $P_T$ reaches its maximum value.}
  \label{figbinfoneRPTchi}
\end{figure}

By numerically solving Eq. (\ref{uk}), we obtain the GW power spectra $P_T^{(s)}$ for various values of the coupling constant $\xi$, and we present the results in Figs. \ref{figbinfoneAPT} and \ref{figbinfoneRPTchi}. Note that the GW power spectrum $P_T$ is bounded from above by the constraint on the tensor-to-scalar ratio, i.e., $r_{0.002}<0.035$ \cite{BICEP:2021xfz}, as indicated by observations at the CMB scale.



From Figs. \ref{figbinfoneAPT} and \ref{figbinfoneRPTchi}, it can be observed that the scale at which the parity-violating effect in primordial GWs (i.e., $\Delta \chi$) reaches its maximum almost coincides with the scale at which the GW power spectrum (i.e., $P_T$) reaches its maximum. Therefore,  for the coupling of the background field with the parity-violating
term in the bounce-inflation model, the scale at which the primordial GW power spectrum is most likely to be observed is also the scale at which the parity-violating effect is most likely to be observed.



\subsection{Coupling the spectator field with the parity-violating term}\label{2.2}


For the coupling of the spectator field with the parity-violating term in the bounce-inflation model, we require that the background evolution is still dominated by $\phi$.
The Lagrangian density of the spectator field $\psi$ is $\mathcal{L}_{\psi}=-\frac{1}{2}\nabla_{\mu}\psi\nabla^{\mu}\psi-U(\psi)$, where $\psi$ is assumed to be the axion.
The axion potential \cite{Kobayashi:2015aaa} is
\be
U\lf(\psi\rt)={1 \over 2}m_a^2{\psi}^2+\Lambda_a^4{\psi \over f_a}\sin\lf({\psi \over f_a}\rt)\,,\label{upsi}
\ee
where the axion mass $m_a=\Lambda_a^2/f_a$, $f_a$ is the axion decay constant, and $\Lambda_a$ is the scale of non-perturbative physics. We require that $\psi$ is minimally coupled with $\phi$. The equation of motion of $\psi$ can be given as
\be \ddot{\psi}+3H\dot{\psi}+\frac{\d U}{\d \psi}=0\,.\label{eom:psi}
\ee
The energy density and pressure of $\psi$ can be given as $\rho_\psi=\dot{\psi}^2/2+U\lf(\psi\rt)$ and $p_\psi=\dot{\psi}^2/2 -U\lf(\psi\rt)$, respectively.
The contribution of $\psi$ to the background evolution should be much smaller than that of $\phi$.

\begin{figure}[htbp]
  \centering
  \hspace{0mm}
  \subfigure[]{\includegraphics[width=0.47\textwidth]{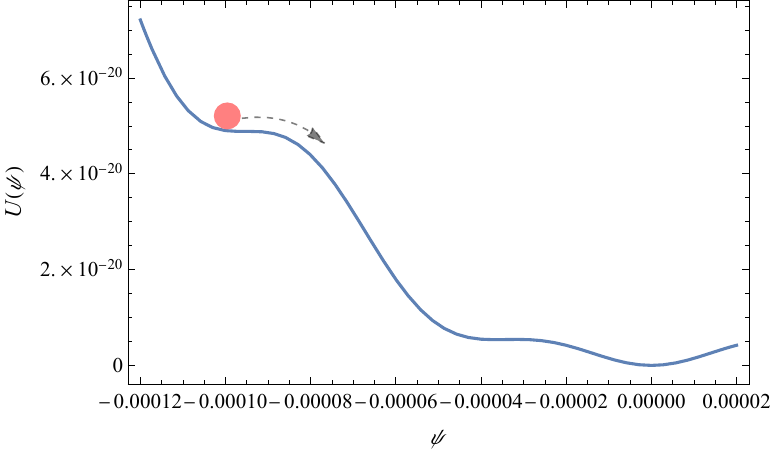}}
  \hspace{0mm}
  \subfigure[]{\includegraphics[width=0.45\textwidth]{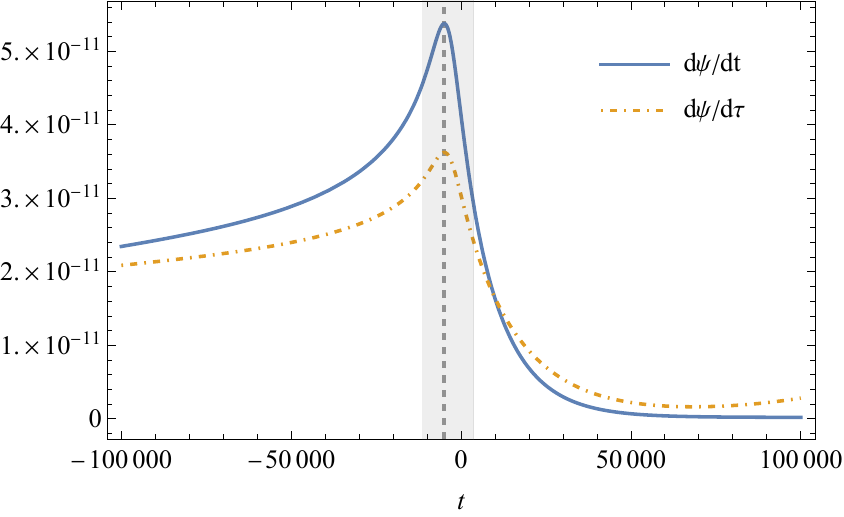}}
  \caption{The potential $U\lf(\psi\rt)$ of the axion $\psi$ and its evolution in our bounce-inflation model for $m_a=3.32\times 10^{-6}$, $f_a=1\times10^{-5}$, $\Lambda_a^4=m_a^2 f_a^2$. The initial conditions of the spectator $\psi$ are set as $\psi({t_i}) = -1\times 10^{-4}$ and $\dot{\psi}({t_i})=3.1\times 10^{-12}$. The NEC-violating phase is depicted by the gray region and the vertical dashed line represents the moment when $H=0$. }
  \label{figbinftwoaxion}
\end{figure}

In this subsection, the spectator field $\psi$ is coupled to the Nieh-Yan term, i.e., $F=\psi$. The amplitude of the primordial GW power spectrum $P_T$ primarily depends on the Hubble parameter $H$ (or equivalently, the evolution of $\phi$), while the strength of the parity-violating effect $\Delta \chi$ mainly relies on the evolution of $\psi$ over time, particularly on its rate $\dot{\psi}$.
The background evolution is still illustrated in Fig. \ref{figbinfbg}. The axion potential $U(\psi)$ and the evolution of $\psi$ are depicted in Fig. \ref{figbinftwoaxion}.

\begin{figure}[htbp]
  \centering
  \hspace{0mm}
  \subfigure[]{\includegraphics[width=0.32\textwidth]{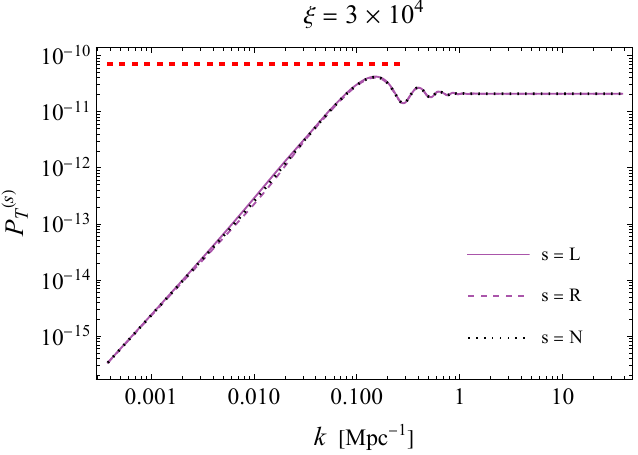}}
  \hspace{0mm}
  \subfigure[]{\includegraphics[width=0.32\textwidth]{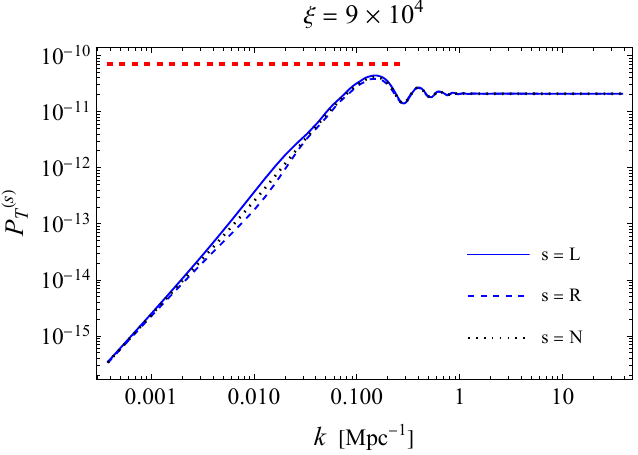}}
  \hspace{0mm}
  \subfigure[]{\includegraphics[width=0.32\textwidth]{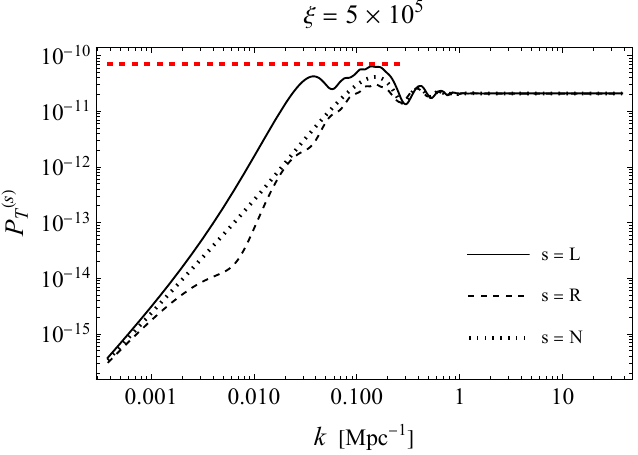}}
  \caption{The primordial GW power spectra $P_T^{(s)}$ for the left-handed ($s=L$, solid curves) and right-handed ($s=R$, dashed curves) modes in the bounce-inflation model for different values of $\xi$ when the spectator $\psi$ is coupled with the Nieh-Yan term. For comparison, we also present the GW power spectra without parity violation ($s=N$, black dotted curves), corresponding to the case when $F \equiv 0$. The red horizontal dashed line denotes the constraint on the tensor-to-scalar ratio, i.e., $r_{0.002}<0.035$.
  }
  \label{figbinftwoAPT}
\end{figure}

\begin{figure}[htp]
  \centering
  \hspace{0mm}
  \subfigure[]{\includegraphics[width=0.43\textwidth]{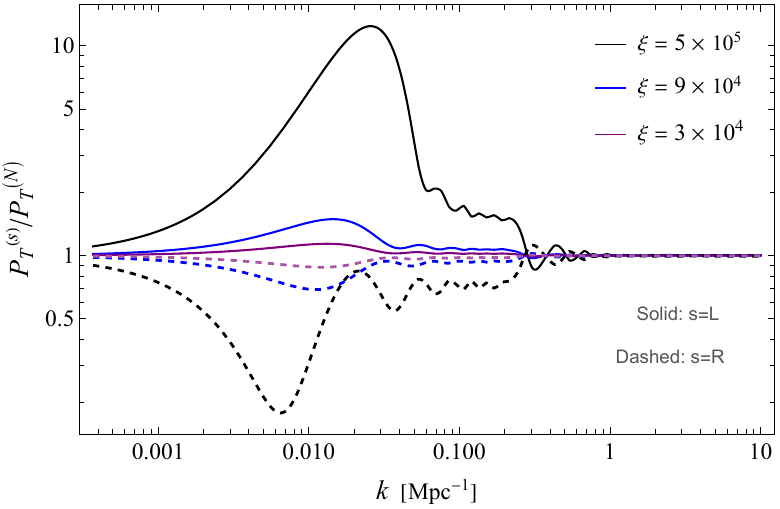}}
  \hspace{0mm}
  \subfigure[]{\includegraphics[width=0.43\textwidth]{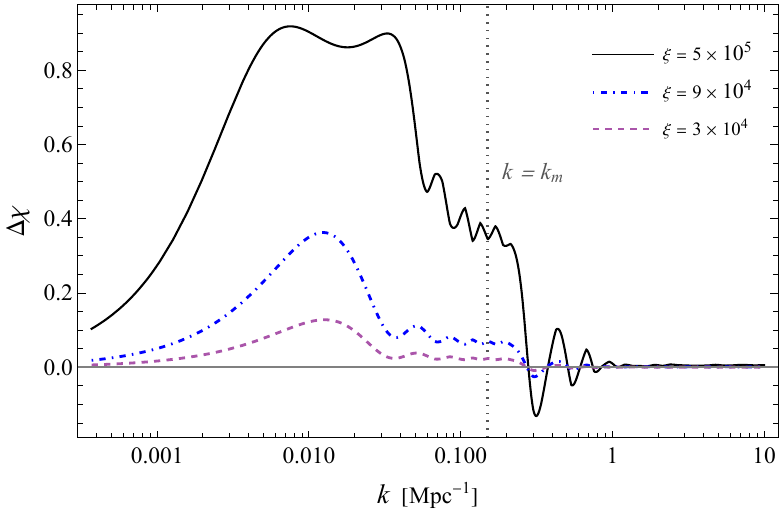}}
  \caption{The ratio of the primordial GW power spectra with and without parity violation ($P_T^{(s)}/P_T^{(N)}$) for different values of $\xi$ in the bounce-inflation model when the spectator $\psi$ is coupled with the Nieh-Yan term, where $s=L$ and $R$ for the solid and dashed curves, respectively. The chiral parameter $\Delta \chi$ is shown in the right panel, where $k=k_m$ corresponds to the wavenumber at which the power spectrum $P_T$ reaches its maximum value.}
  \label{figbinftwoRPTchi}
\end{figure}

Similar to Sec. \ref{2.1}, we obtain $P_T^{(s)}$, $P_T^{(s)}/P_T^{(N)}$ and $\Delta\chi$ by numerically solving Eq. (\ref{uk}). The corresponding results are presented in Figs. \ref{figbinftwoAPT} and \ref{figbinftwoRPTchi}.
Notably, the scale at which the parity-violating effect in primordial GWs (i.e., $\Delta \chi$) reaches its maximum is significantly misaligned with the scale at which the GW power spectrum (i.e., $P_T$) reaches its maximum. The relatively large parity-violating effect occurs at scales larger than those corresponding to the maximum of $P_T$. Since $P_T$ is suppressed at large scales, the corresponding parity-violating effect is very unlikely to be observed. This is significantly different from the case where the background field couples with the Nieh-Yan term in the bounce-inflation model. We will analyze the reason for this difference in Sec. \ref{conclusion}.
It can be inferred that a similar conclusion can be drawn for the Genesis-inflation scenario.

\section{Parity-violating primordial GWs from NEC violation during inflation}\label{3}

In this section, we investigate the parity violation effect of primordial GWs in a scenario where an intermediate NEC violation occurs between two stages of slow-roll inflation \cite{Cai:2020qpu,Cai:2022nqv}. In this scenario, the energy scale of the second stage of inflation is significantly higher than that of the first stage.
Consequently, this scenario can produce a significantly enhanced power spectrum of primordial GWs at small scales (such as the PTA scale) while remaining consistent with observational constraints at CMB scale. Such models have the potential to explain the latest PTA observational data \cite{Ye:2023tpz} (see also \cite{Jiang:2023gfe}). The parity violation effect was investigated in this scenario for the case where the background field is coupled with the parity-violating term \cite{Cai:2022lec}.
Similar to Sec. \ref{section2}, we investigate the differences in the parity-violating effects produced when the background field and the spectator field are coupled with the Nieh-Yan term, respectively.

We work with the effective action
\be
S=\int {\rm d}^4x\sqrt{-g}\lf[{M_{\text{P}}^2 \over 2}R-M_{\text{P}}^2{g_1\lf(\phi\rt) \over 2}X+{g_2\lf(\phi\rt) \over 4}X^2-M_{\text{P}}^4V\lf(\phi\rt)+\mathcal{L}_{NY}+\mathcal{L}_{\psi}\rt]\,,\label{action:NECV240530}
\ee
where
\be
g_1\lf(\phi\rt)=\frac{2}{1+e^{-q_1\lf(\phi-\phi_0\rt)}}
+\frac{1}{1+e^{q_2\lf(\phi-\phi_3\rt)}}
-\frac{f_1e^{2\phi}}{1+f_1e^{2\phi}},
\ee
\be
g_2\lf(\phi\rt)=\frac{f_2}{1+e^{-q_2\lf(\phi-\phi_3\rt)}}
\frac{1}{1+e^{q_3\lf(\phi-\phi_0\rt)}},
\ee
\be
V\lf(\phi\rt)={1 \over 2}m^2\phi^2\frac{1}{1+e^{q_2\lf(\phi-\phi_2\rt)}}
+\lambda\lf[1-\frac{\lf(\phi-\phi_2\rt)^2}{\sigma^2}\rt]^2\frac{1}{1+e^{-q_4\lf(\phi-\phi_1\rt)}},
\ee
$q_{1,2,3,4}$, $f_{1,2}$, $m$ and $\lambda$ are positive constants, see \cite{Cai:2022lec} for details. The scalar field $\phi$ determines the background evolution, while the axion $\psi$  serves as the spectator field.
Again, $\mathcal{L}_{\psi}=-\nabla_{\mu}\psi\nabla^{\mu}\psi/2-U(\psi)$, $\mathcal{L}_{NY}$ represents the coupling of $\phi$ or $\psi$ with the Nieh-Yan term, i.e.,
\be
\mathcal{L}_{NY}={\xi F(\phi,\psi)\over 4}\mathcal{T}_{A\mu\nu}\tilde{\mathcal{T}}^{A\mu\nu}\,,\label{eq:coupling001}
\ee
where $F=\phi$ when the background field is coupled with the Nieh-Yan term and $F=\psi$ when the spectator is coupled with the Nieh-Yan term.

The background equations can be given as
\ba
3H^2M_{\text{P}}^2&=&{M_{\text{P}}^2 \over 2}g_1\dot{\phi}^2+
{3 \over 4}g_2\dot{\phi}^4+M_{\text{P}}^4V+\rho_{\psi}\, ,\\
\dot{H}M_{\text{P}}^2&=&-{M_{\text{P}}^2 \over 2}g_1\dot{\phi}^2
-{1 \over 2}g_2\dot{\phi}^4 - {1\over2}(\rho_\psi+p_\psi) \,,
\label{necbg}
\ea
where the energy density and pressure of the spectator field are $\rho_\psi=\dot{\psi}^2/2+U\lf(\psi\rt)$ and $p_\psi=\dot{\psi}^2/2 -U\lf(\psi\rt)$, respectively.
We require that $\rho_\psi$ and $p_\psi$ are subdominant in the background equations.

\begin{figure}[htpb]
  \centering
  \hspace{0mm}
  \subfigure[]{\includegraphics[width=0.43\textwidth]{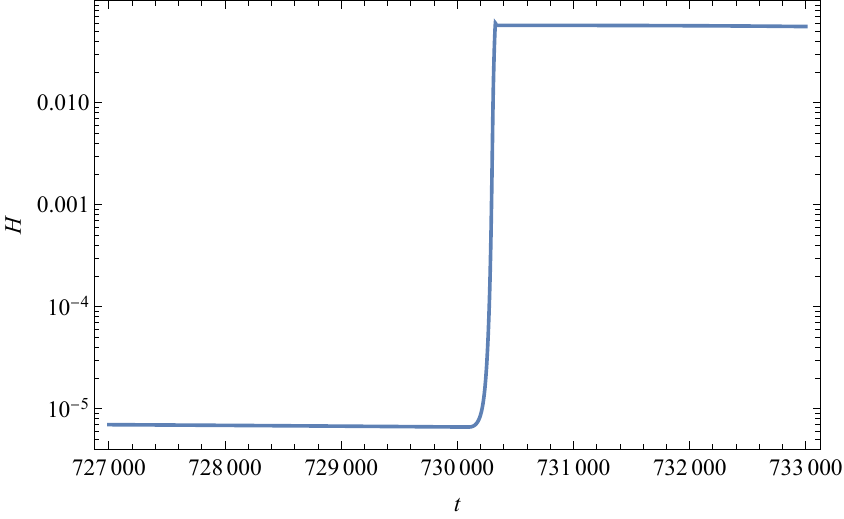}}
  \hspace{0mm}
  \subfigure[]{\includegraphics[width=0.425\textwidth]{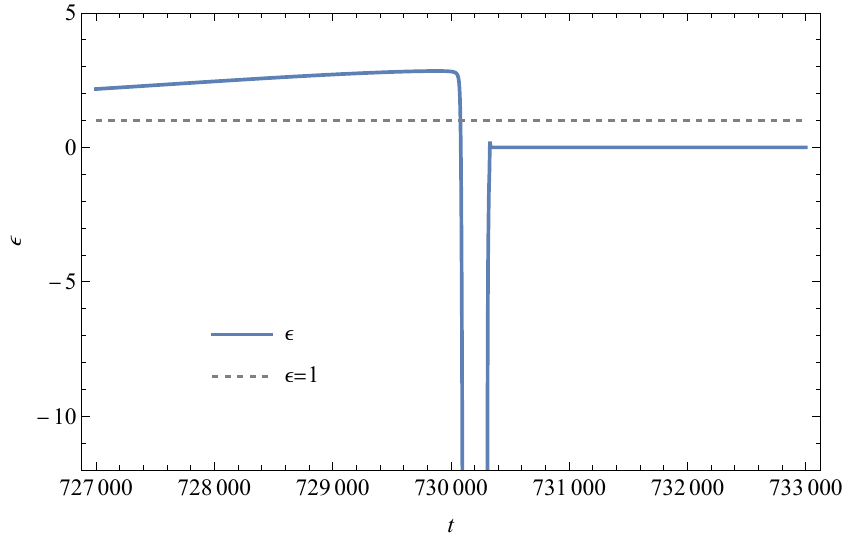}}
  \hspace{0mm}
  \subfigure[]{\includegraphics[width=0.44\textwidth]{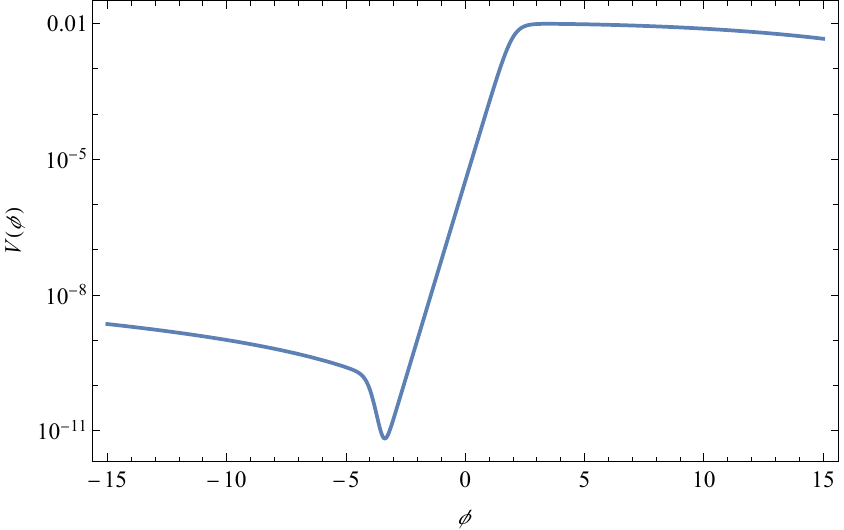}}
  \hspace{0mm}
  \subfigure[]{\includegraphics[width=0.425\textwidth]{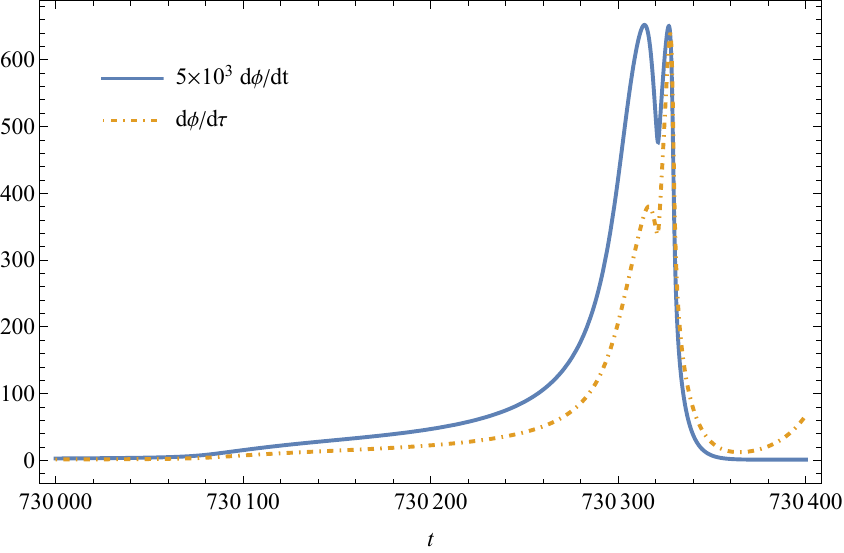}}
  \caption{The background evolution of the scenario involves an intermediate NEC violation occurring between two stages of slow-roll inflation. We have set $\phi_0=3.2$, $\phi_1=2$, $\phi_2=-4$, $\phi_3=-4.38$, $q_1=10$, $q_2=6$, $q_3=10$, $q_4=4$, $f_1=1$, $f_2=40$, $\lambda_1=0.01$, $\sigma=23$, $m=-4.5\times 10^{-6}$, and the initial values of $\phi$ $\phi_i=-7$ and $\dot{\phi}_i=0$.}
  \label{fignecbg}
\end{figure}

The numerical solution results of the background evolution are presented in Fig. \ref{fignecbg}. As we can see, the NEC violation ($\dot{H}>0$) during inflation significantly disrupts the slow-roll condition. The background field $\phi$ can climb up a steep potential barrier, and then enter a second phase of slow-roll inflation with a significantly higher energy scale (i.e., $H_{\text{inf}_2}\gg H_{\text{inf}_1}$), see \cite{Cai:2020qpu} for details.
Since $\psi$ remains a spectator field, the evolution of the background in the two different coupling scenarios is approximately equivalent.

In the following, we will numerically obtain the power spectra of the primordial GWs in the two different coupling scenarios. Note that Eqs. (\ref{hk}) to (\ref{eq:Chi}) remain applicable for the action (\ref{action:NECV240530}).

\subsection{Coupling the background field with the parity-violating term}\label{3.1}

In this subsection, we numerically solve Eq. (\ref{hk}) and present the results for the power spectra (\ref{eq:power240601}) of parity-violating primordial GWs and the parity violation parameter $\Delta\chi$ within a model that exhibits intermediate NEC violation during inflation. We set $\mathcal{L}_{\psi}=0$ and $F=\phi$ in this subsection.

In this case, the power spectrum of parity-violating primordial GWs has been calculated in \cite{Cai:2022lec}. The results indicate that NEC violation during inflation can significantly enhance the power spectrum of primordial GWs at higher frequencies and substantially increase the observability of the parity-violating effect. However, it is worth noting that the parity-violating effect obtained in \cite{Cai:2022lec} remains too small, with the maximum value of $\Delta \chi$ only at a few percent level. This is because the parity-violating term used in \cite{Cai:2022lec} is the Chern-Simons term. To avoid ghost instability within the frequency range of interest, $\Delta \chi$ is strongly constrained.

\begin{figure}[htpb]
  \centering
  \hspace{0mm}
  \subfigure[]{\includegraphics[width=0.32\textwidth]{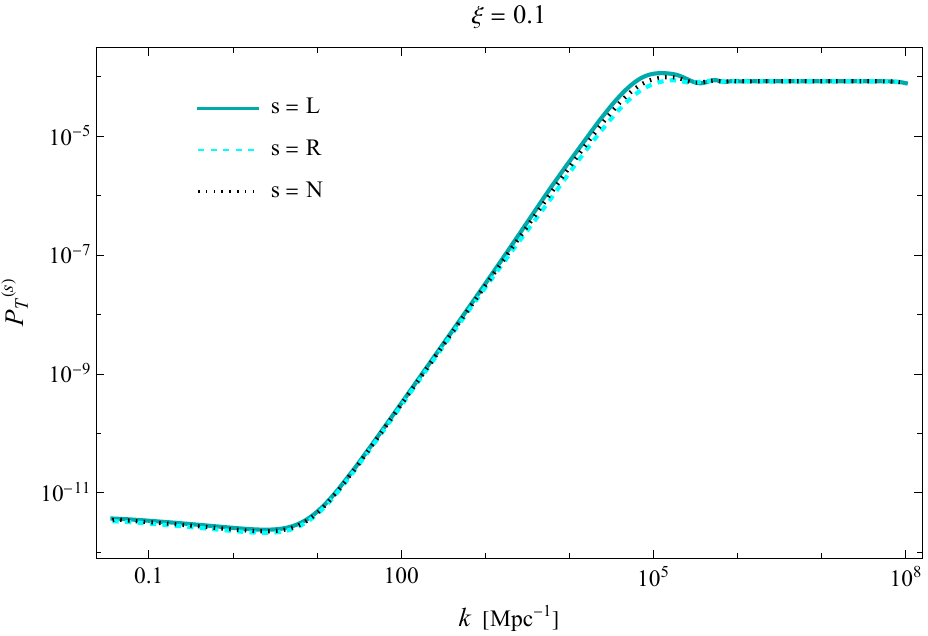}}
  \hspace{0mm}
  \subfigure[]{\includegraphics[width=0.32\textwidth]{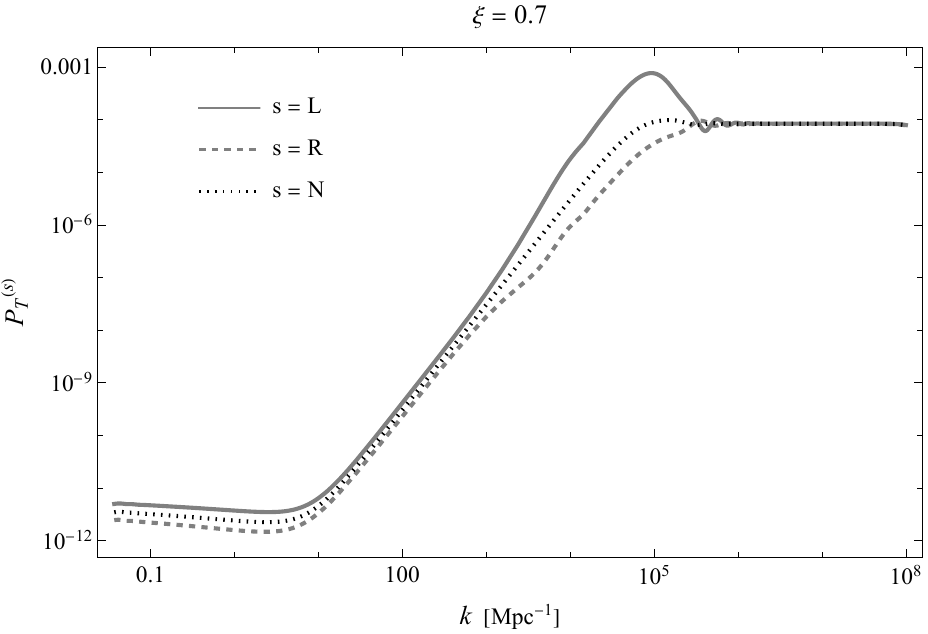}}
  \hspace{0mm}
  \subfigure[]{\includegraphics[width=0.32\textwidth]{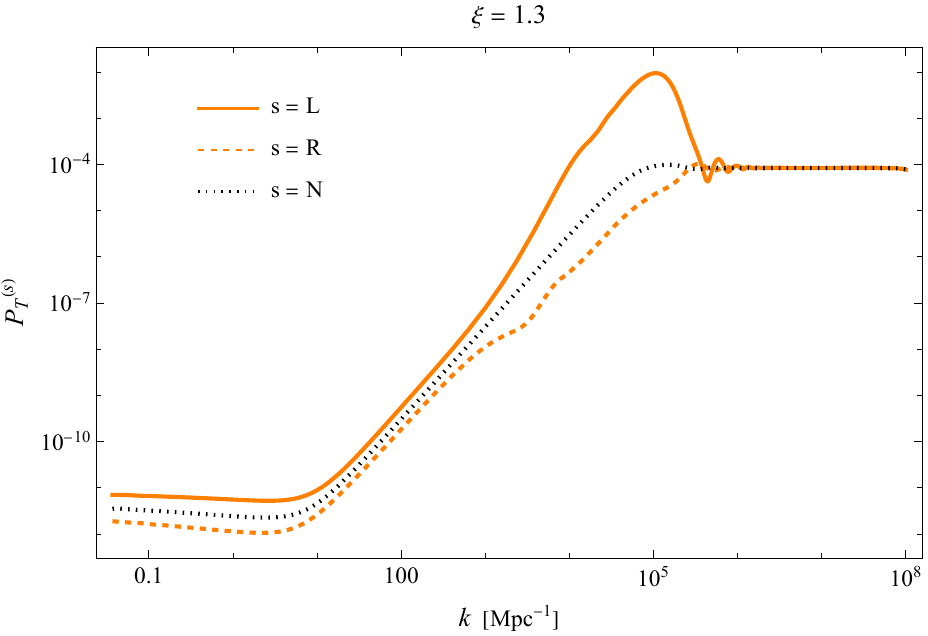}}
  \hspace{0mm}
  \caption{The primordial GW power spectra $P_T^{(s)}$ for the left-handed ($s=L$, solid curves) and right-handed ($s=R$, dashed curves) modes in the model that exhibits intermediate NEC violation during inflation for different values of $\xi$ when the background field $\phi$ is coupled with the Nieh-Yan term. For comparison, we also present the GW power spectra without parity violation ($s=N$, black dotted curves), corresponding to the case when $F \equiv 0$.}
  \label{figneconeApt}
\end{figure}

\begin{figure}[htpb]
  \centering
  \hspace{0mm}
  \subfigure[]{\includegraphics[width=0.43\textwidth]{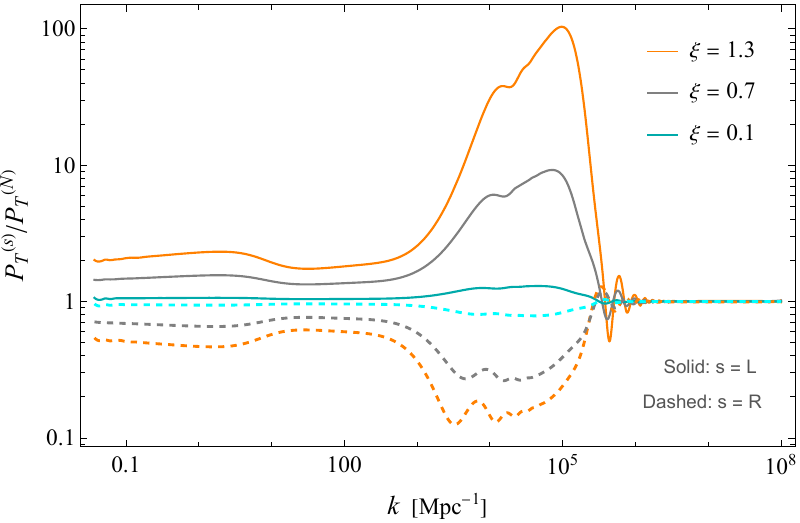}}
  \hspace{0mm}
  \subfigure[]{\includegraphics[width=0.435\textwidth]{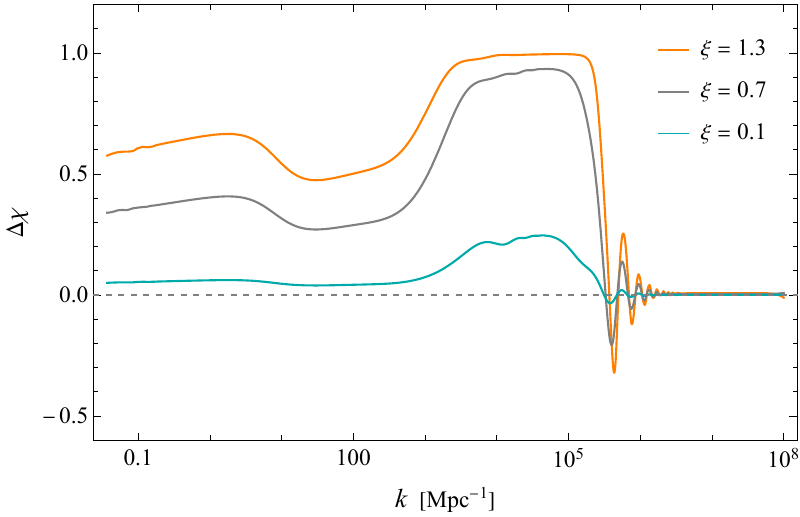}}
  \hspace{0mm}
  \caption{The ratio of the primordial GW power spectra with and without parity violation ($P_T^{(s)}/P_T^{(N)}$) for different values of $\xi$ in the model that exhibits intermediate NEC violation during inflation when the background field $\phi$ is coupled with the Nieh-Yan term, where $s=L$ and $R$ for the solid and dashed curves, respectively. The chiral parameter $\Delta \chi$ is shown in the right panel.}
  \label{figneconeRpt}
\end{figure}

In this paper, we replace the Chern-Simons term in \cite{Cai:2022lec} with the Nieh-Yan term, allowing us to obtain a sufficiently large $\Delta \chi$.
As shown in Figs. \ref{figneconeApt} and \ref{figneconeRpt}, the frequency at which the parity-violating effect in the primordial GWs is most pronounced is almost the same as the frequency at which the power spectrum of the primordial GWs reaches its maximum.
As shown in \cite{Ye:2023tpz}, the primordial GW power spectrum generated by our model can explain the latest PTA observational data \cite{NANOGrav:2023hvm,NANOGrav:2023gor,EPTA:2023fyk,Reardon:2023gzh,Xu:2023wog}. Therefore, the parity-violating primordial GW power spectrum in Figs. \ref{figneconeApt} and \ref{figneconeRpt} may be observationally interesting.

\subsection{Coupling the spectator field with the parity-violating term} \label{3.2}

In this subsection, we numerically solve Eq. (\ref{hk}) and obtain the power spectra of parity-violating primordial GWs and the parity violation parameter $\Delta\chi$ in a model that exhibits intermediate NEC violation during inflation when the spectator $\psi$ is coupled with the Nieh-Yan term. The background evolution is driven by $\phi$. Therefore, the background evolution can still be represented by Fig. \ref{fignecbg}.

\begin{figure}[htpb]
  \centering
  \hspace{0mm}
  \subfigure[]{\includegraphics[width=0.43\textwidth]{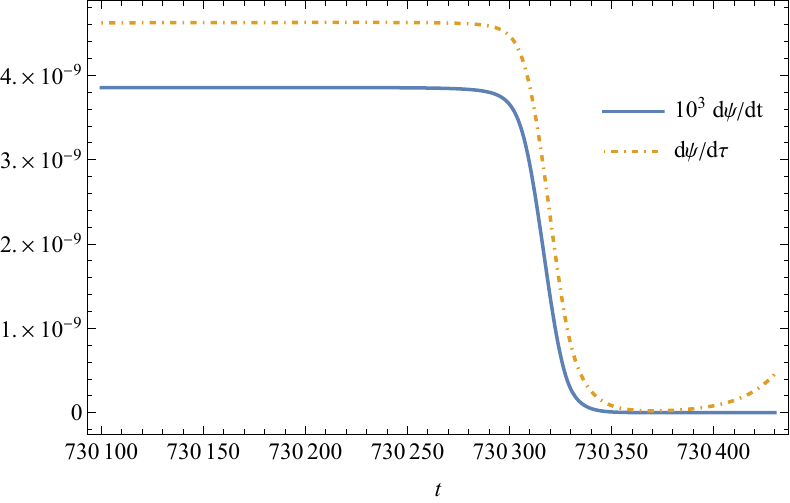}}
  \hspace{0mm}
  \subfigure[]{\includegraphics[width=0.43\textwidth]{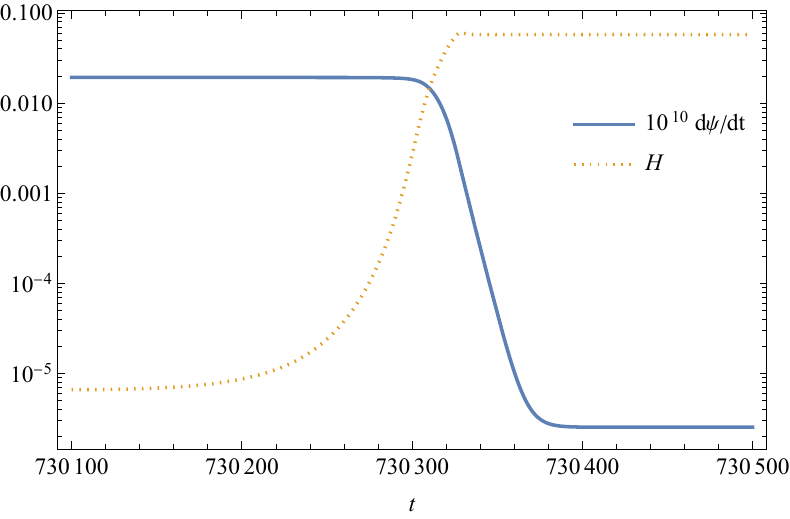}}
  \caption{The evolution of $\dot{\psi}$ and comparison with the Hubble parameter $H$ for $m=3.32\times10^{-6}$, $f_a=1\times10^{-5}$, $\Lambda_a^4=m^2 f_a^2$, $\psi_i = -1\times 10^{-4}$, and $\dot{\psi}_i=3.1\times 10^{-12}$. }
  \label{fignecaxion}
\end{figure}

Similar to Sec. \ref{2.2}, we couple $\psi$ to the Nieh-Yan term as given in Eq. (\ref{eq:coupling001}), where $F=\psi$ and $\mathcal{L}_{\psi}=-\nabla_{\mu}\psi\nabla^{\mu}\psi/2-U(\psi)$. The axion potential $U(\psi)$ and the equation of motion for $\psi$ are still given by Eq. (\ref{upsi}) and (\ref{eom:psi}), respectively. For the potential displayed in Fig. \ref{figbinftwoaxion}(a), we provide the evolution of $\dot{\psi}$ in Fig. \ref{fignecaxion}, where the initial conditions are set the same as in Fig. \ref{figbinftwoaxion}. Due to the NEC violation, the Hubble parameter $H$ increases rapidly by several orders of magnitude. As a result, the friction term in Eq. (\ref{eom:psi}) increases rapidly, causing $\dot{\psi}$ to decrease sharply near the end of the NEC violation.

\begin{figure}[htpb]
  \centering
  \subfigure[]{\includegraphics[width=0.32\textwidth]{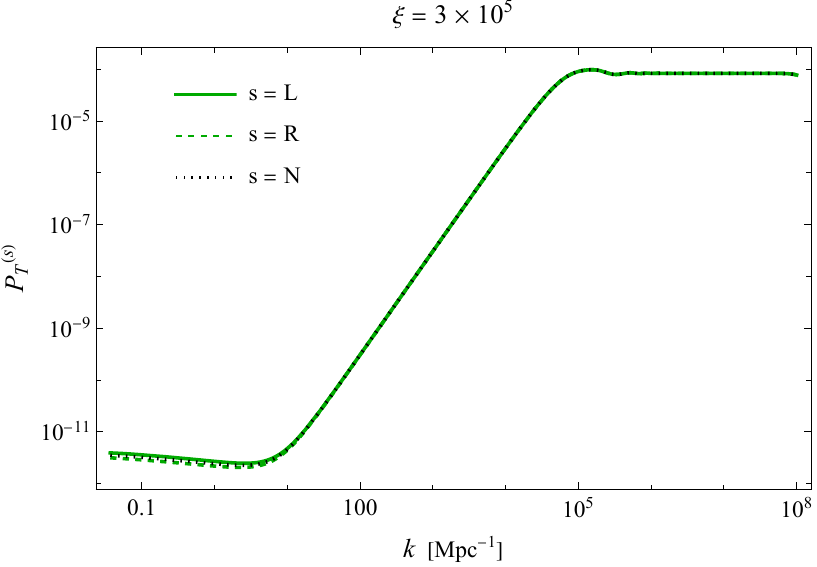}}
  \hspace{0mm}
  \subfigure[]{\includegraphics[width=0.32\textwidth]{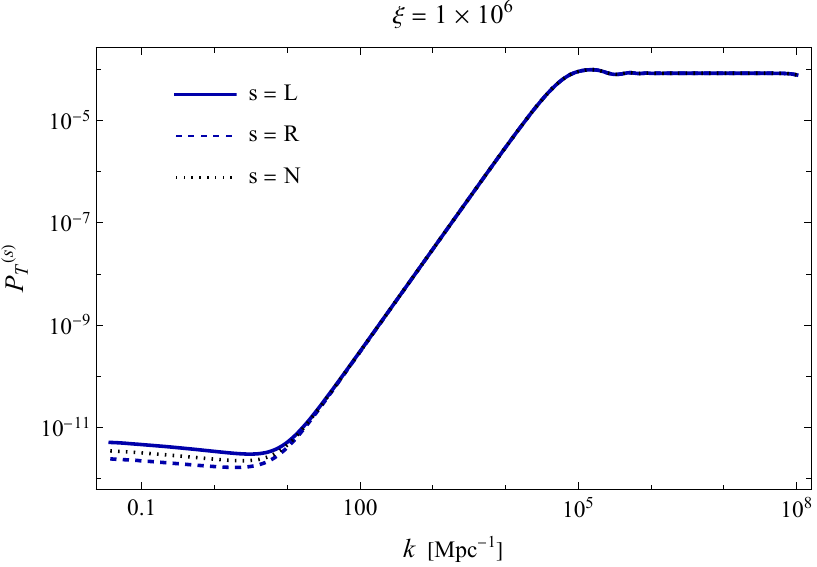}}
  \hspace{0mm}
  \subfigure[]{\includegraphics[width=0.32\textwidth]{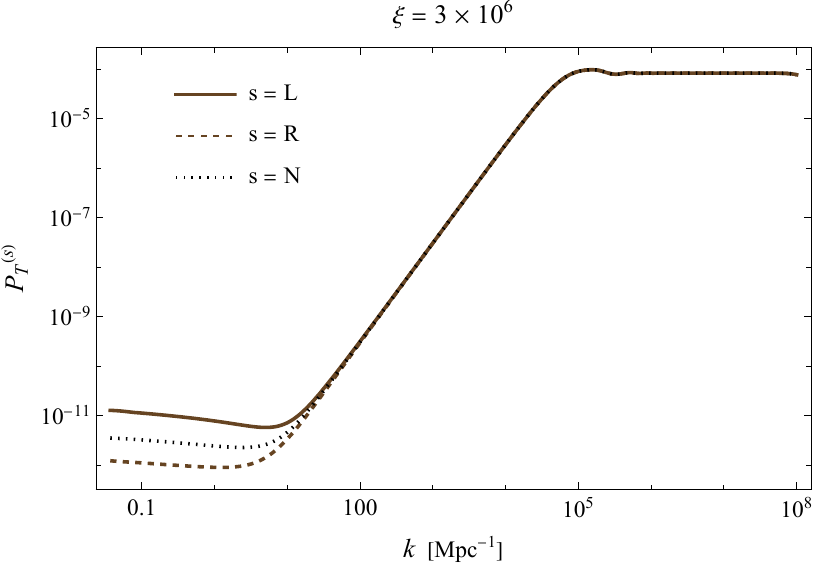}}
  \caption{The primordial GW power spectra $P_T^{(s)}$ for the left-handed ($s=L$, solid curves) and right-handed ($s=R$, dashed curves) modes in the model that exhibits intermediate NEC violation during inflation for different values of $\xi$, where the spectator $\psi$ is coupled with the Nieh-Yan term. For comparison, we also present the GW power spectra without parity violation ($s=N$, black dotted curves), corresponding to the case when $F \equiv 0$.}
  \label{fignectwoApt}
\end{figure}

\begin{figure}[htp]
  \centering
  \hspace{0mm}
  \subfigure[]{\includegraphics[width=0.43\textwidth]{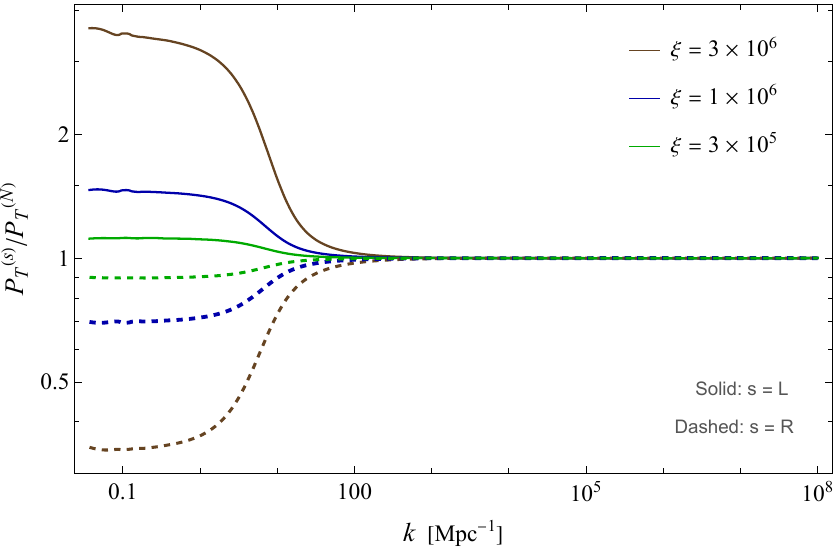}}
  \hspace{0mm}
  \subfigure[]{\includegraphics[width=0.43\textwidth]{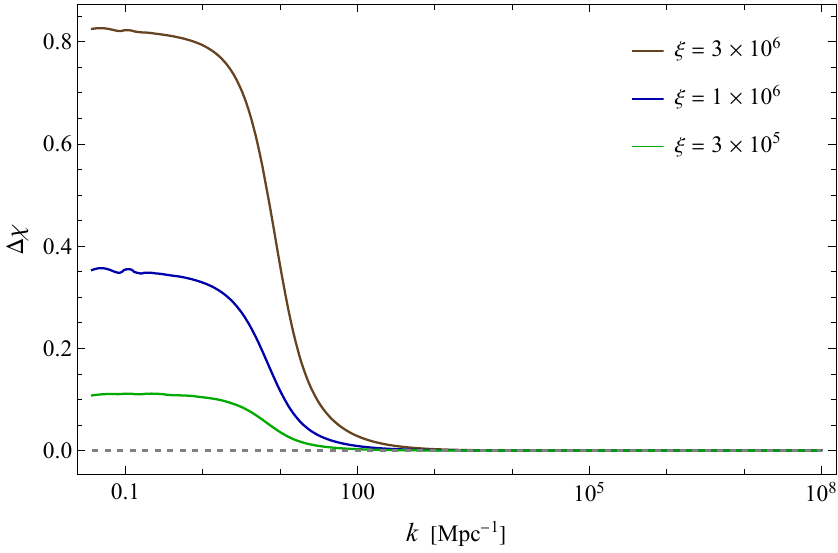}}
  \caption{The ratio of the primordial GW power spectra with and without parity violation ($P_T^{(s)}/P_T^{(N)}$) for different values of $\xi$ in the model that exhibits intermediate NEC violation during inflation when the spectator $\psi$ is coupled with the Nieh-Yan term, where $s=L$ and $R$ for the solid and dashed curves, respectively. The chiral parameter $\Delta \chi$ is shown in the right panel. }
  \label{fignectwoRPT}
\end{figure}

We numerically obtain the primordial GW power spectra for different polarization modes and the parity violation parameter, as shown in Figs. \ref{fignectwoApt} and \ref{fignectwoRPT}. Similar to the results in Sec. \ref{2.2}, significant parity-violating effects are prominent in the low-frequency range, where the power spectrum of primordial GWs has not been significantly enhanced. This indicates that observing corresponding parity-violating effects in the high-frequency range (such as the observational windows of PTA and laser interferometers) is unpromising.


Although modifying the initial conditions might allow $\dot{\psi}$ to reach a larger maximum value, as a spectator field, its behavior near the end of the NEC violation and afterward is primarily governed by the friction term. Additionally, for reasons discussed in Sec. \ref{sec:why-spectator}, the model in this subsection can only produce significant parity violation in the relatively low-frequency range, where the power spectrum remains small, and it is challenging to achieve significant parity violation in the high-frequency range. Therefore, for models with NEC violation during inflation, it is difficult for spectators to produce significant observable parity violation effects near the NEC violation unless $\dot{\psi}$ is deliberately designed to reach a large value during the second stage of slow-roll inflation.

\section{Discussions}\label{Sec:Discuss}

In this section, we briefly discuss the reasons why NEC violation can produce blue-tilted power spectra and why spectators cannot generate observationally significant parity violation during NEC violation.

\subsection{Blue-tilted power spectra from NEC violation}\label{sec-blue}

Blue-tilted power spectra from NEC violation have been investigated in \cite{Cai:2015yza,Cai:2016ldn,Cai:2020qpu,Cai:2022nqv,Cai:2022lec,Zhu:2023lbf,Ye:2023tpz,Pan:2024ydt}. Here, for simplicity, we consider parity-conserving primordial GWs with a propagating speed equal to the speed of light (i.e., unity). The perturbation equation can be written as follows:
\be
{u_k}''+\lf(k^2-{a'' \over a}\rt){u_k }=0 \,, \label{eq:eomu0610}
\ee
where $u_k\equiv a h_k$, $h_k$ is the primordial tensor perturbations mode.

It is well-known that on superhorizon scales (i.e., $k^2\ll a''/a$), the solution to Eq. (\ref{eq:eomu0610}) can be decomposed into a constant mode and an evolving mode, i.e.,
\be
h_k = h_{\text{const}} + h_{\text{evol}}\,,\label{eq:hk0816}
\ee
where
\be
h_{\text{const}} = C_k\,,\qquad
h_{\text{evol}} = D_k\int a^{-2}\d \tau \,,
\ee
$C_k$ and $D_k$ are $k$-dependent constants.
For simplicity, we parameterize the scale factor as
\be
a \sim t^{\frac{2}{3(1+w)}} \sim \tau^{\frac{2}{1+3w}} \,,
\ee
where $w$ represents the equation of state parameter and is assumed to be constant during a certain stage.

During slow-roll inflation, we approximately have $w=-1$ and $a\sim |\tau|^{-1}$, which results in $h_{\text{evol}}\sim |\tau|^3$, where $\tau<0$. Therefore, during inflation, the evolving mode of primordial GWs, i.e., $h_{\text{evol}}$, decreases on super-horizon scales as the universe expands.
During the NEC violation, where $w<-1$, we can obtain $h_{\text{evol}}\sim |\tau|^{1-{4\over 1+3w}}$, which also decreases on super-horizon scales as the universe expands.
This behavior is different from the evolving mode of the primordial curvature perturbation, which grows on super-horizon scales during NEC violation (see the Supplemental Material of \cite{Cai:2023uhc}), thereby leading to a prominent peak in the power spectrum of primordial curvature perturbations. It is such a peak that can lead to the production of primordial black holes with masses and abundances of observational interest \cite{Cai:2023uhc}.

The decrease of $h_{\text{evol}}$ indicates that the perturbation mode $h_k$ is dominated by the constant mode $h_{\text{const}}$ shortly after it exits the horizon. Since $h_{\text{const}}$ primarily depends on the background evolution (i.e., $H$) when the perturbation mode exits the horizon, the power spectrum of perturbation modes that exit the horizon during inflation is nearly scale-invariant, whereas the power spectrum of perturbation modes that exit the horizon during NEC violation is blue-tilted.

The spectral index of the blue spectrum can be estimated as follows.
When the perturbation mode $h_k$ with wavenumber $k$ is within the horizon, we have $u_k \approx e^{-ik\tau}/\sqrt{2k}$. When $h_k$ exits the horizon during NEC violation, we still approximately have $|u_k| \simeq 1/\sqrt{2k}$ and $k^2 \simeq a''/a \sim |\tau|^{-2}$. Based on the aforementioned parametrization of the scale factor, we can obtain
\be a \sim k^{-2/(1+3w)} \sim H^{-2/[3(1+w)]} \label{eq:akH01}
\ee
at the moment of horizon crossing.

More precisely, the relationship between $k$ and $a$ at the time of horizon crossing should be obtained through the equation
\be \int_{t_i}^t \frac{1}{a(\tilde{t})} d\tilde{t} \simeq \frac{2\pi}{k} \,,\label{horizon001}
\ee
where the integral on the left-hand side represents the comoving distance traveled by perturbation modes propagating at the speed of light from $t_i$ to $t$, known as the comoving particle horizon of the perturbation modes. This still yields the relation presented in Eq. (\ref{eq:akH01}).

Since the perturbation mode $h_k$ is dominated by the constant mode after exiting the horizon, the power spectrum can be estimated by evaluating $|h_k|$ at the time of horizon crossing.
Thus, we have
\be |h_k| \equiv |u_k|/a \sim H^{2/[3(1+w)]}/\sqrt{2k} \ee
at the time of horizon crossing. As a result, we find that the power spectrum can be written as
\be P_T \equiv \frac{k^3}{2\pi^2}\lf|h_k\rt|^2 \sim k^{n_T}
\ee
for the perturbation modes exiting the horizon during the NEC-violating phase, where the spectral index
\be n_T = 2 + {4/(1+3w)} \,,
\ee
$w<-1$ is the equation of state parameter of the NEC-violating phase.
This result is consistent with that obtained in \cite{Cai:2022nqv}.
Evidently, in the limit of $w \rightarrow -1$, $n_T \rightarrow 0$, which approaches the slow-roll inflation scenario. Since $w < -1$ during NEC violation, we have $n_T > 0$, indicating a blue spectrum.

\subsection{Why cannot spectators generate observationally significant parity violation during NEC violation?}
\label{sec:why-spectator}

After introducing the parity-violating term, Eq. (\ref{eq:eomu0610}) is modified to Eq. (\ref{uk}), where $c_T^{(s)}$ in Eq. (\ref{uk}) can be regarded as the effective sound speed for perturbation modes of different polarizations. Consequently, Eq. (\ref{horizon001}) changes to
\be \int_{t_i}^t \frac{c_T^{(s)}(\tilde{t})}{a(\tilde{t})} d\tilde{t} \simeq \frac{2\pi}{k} \,,\label{horizon002}
\ee
where ${c_T^{(s)}}\equiv \sqrt{1+\mu^{(s)}}$. Evidently, perturbation modes of different polarization that exit the horizon at the same time have different comoving wavenumbers (or comoving wavelengths), and the larger the value of $\mu^{(s)}$, the greater this difference.

Furthermore, for an effective sound speed for perturbations that evolves non-trivially over time, the $k$ or $a$ used in the analysis in Sec. \ref{sec-blue} should also be correspondingly modified. This could introduce interesting features in the power spectra of perturbation modes of different polarizations. Similar to the case of scalar perturbations \cite{Cai:2023uhc}, around the NEC violation, the evolving modes $h_{\text{evol}}$ at certain scales outside the horizon may grow and surpass the constant mode.
Perturbation modes with certain wavenumbers will experience more growth than other modes.
Therefore, in the parity-conserving case, the profile of the primordial GW power spectrum in the frequency domain generally does not exhibit peaks or valleys significantly different from the profile of $H^2$ in the time domain. However, the parity-violating primordial GW power spectrum may exhibit significant peaks or valleys, making it notably different from the profile of $H^2$. The evolution behavior of perturbation modes near NEC violation is very complex, and we will not delve into it further here.

Roughly speaking, the perturbation modes that exit the horizon when $|\mu^{(s)}|$ reaches its maximum exhibit the greatest parity-violating effect in the power spectrum. This is reflected in Figs. \ref{figbinfoneAPT}, \ref{figbinftwoAPT}, \ref{figneconeApt} and \ref{fignectwoApt} as the maximum deviation of the power spectrum compared to the power spectrum without parity violation.
Based on the convention in this paper, with $F'>0$, the integral on the left side of Eq. (\ref{horizon002}) for the right-handed mode ($\lambda^{(R)}=+1$) is larger than that in the parity-conserving case ($\lambda^{(N)}=0$), while the integral for the left-handed mode ($\lambda^{(L)}=-1$) is smaller than that in the parity-conserving case ($\lambda^{(N)}=0$).
Therefore, with the same parameter settings of the model, the wavenumber corresponding to the minimum of $P_T^{(R)}/P_T^{(N)}$ is smaller than the maximum of $P_T^{(L)}/P_T^{(N)}$, as shown in Figs. \ref{figbinfoneAPT}, \ref{figbinftwoAPT}, \ref{figneconeApt} and \ref{fignectwoApt}.

In the case where the background field $\phi$ is coupled with the parity-violating term, we roughly have $\lambda^{(s)}\xi F'/k_{\text{max}}^{(s)} \lesssim 1$ (or $\lambda^{(s)}\xi F'/k_{\text{max}}^{(s)}$ is at most of the order of 10 for some of the values of $\xi$ we consider in this paper), in which $k_{\text{max}}^{(s)}$ is given by Eq. (\ref{horizon002}) at $t=t_{\text{max}}$, with $t_{\text{max}}$ being the time when $F'$ reaches its maximum value. This implies that the $k_{\text{max}}^{(s)}$ obtained from solving Eq. (\ref{horizon002}) will not differ in order of magnitude from the $k_{\text{max}}^{(s)}$ obtained from solving Eq. (\ref{horizon001}) (i.e., when $\xi=0$). Furthermore, since the times when $F'$ and $H$ reach their maximum values are very close, the wavenumber corresponding to the maximum parity violation in the case where the background field is coupled with the parity-violating term is approximately the same as the wavenumber corresponding to the maximum value of the power spectrum (see the numerical results in the previous sections).

In the case where the spectator $\psi$ is coupled with the parity-violating term, we have $|\dot{\psi}| \ll |\dot{\phi}|$. Even if a larger $\xi$ is chosen such that the maximum value of $\xi \dot{\psi}$ reaches a level comparable to the maximum value of $\xi \dot{\phi}$, the resulting parity-violating effect is still unpromising to be observable on PTA or smaller scales.
This can be understood as follows. The spectator $\psi$ does not have the acceleration mechanism that $\phi$ has,\footnote{In fact, when $H>0$, the friction term $3H\dot{\psi}$ always acts as a hindrance to the increase of $\dot{\psi}$, and this hindrance becomes more pronounced as $H$ increases. During the contraction phase of bouncing inflation, when $H<0$, the friction term acts as an accelerator, but once entering the expansion phase, it also acts as a hindrance.} which means that if $\xi \dot{\psi}$ reaches a level comparable to the maximum value of $\xi \dot{\phi}$, then the integral on the left side of Eq. (\ref{horizon002}) will be significantly larger than that in the case where the background field is coupled with the parity-violating term  (i.e., the left side of Eq. (\ref{horizon001})), and the corresponding $k_{\text{max}}^{(s)}$ will be much smaller (in fact, $\lambda^{(s)}\xi F'/k_{\text{max}}^{(s)} \gg 1$ in this case).
Therefore, as we see from the numerical results in Sec. \ref{3.2}, the wavenumber corresponding to the maximum parity-violating effect is more than an order of magnitude smaller than the wavenumber corresponding to the maximum value of the power spectrum (specifically depending on the value of $\xi$).

As a result, spectator fields cannot benefit from the enhancement of the observability of parity-violating effects due to NEC violation as the background field does. This conclusion will not be significantly affected by modifications to $V(\phi)$ and $U(\psi)$, as long as the role of $\psi$ as a spectator field remains unchanged. Although it may seem that a sufficiently large $\xi$ could cause a more significant deviation in the left-handed perturbation modes at the scale corresponding to the maximum value of the power spectrum (see e.g., Fig. \ref{figbinftwoAPT}(c)), an excessively large $\xi\dot{F}$ may cause significant growth of perturbation modes crossing the horizon near the NEC violation, at sub-horizon or near-horizon scales. There could be two types of potential instabilities on certain scales: one is a gradient instability caused by ${c_T^{(s)}}^2<0$, and the other is a tachyonic instability caused by ${c_T^{(s)}}^2k^2 - a''/a < 0$.\footnote{It appears that both ${c_T^{(s)}}^2 < 0$ and ${c_T^{(s)}}^2k^2 - a''/a < 0$ occur in the infrared limit, where $\left|{c_T^{(s)}}^2\right|k^2 \ll a''/a$ (i.e., on super-horizon scales). However, in this situation, the general solution to Eq. (\ref{uk}), i.e., Eq. (\ref{eq:hk0816}), is dominated by the constant mode. Therefore, the infrared region is unaffected by instabilities in the form of GW overproduction.} Therefore, in principle, there should be an upper limit for $\Delta \chi$ to avoid instability and large backreaction. The relevant details require further investigation. Quantitatively determining this upper limit for $\Delta \chi$ is beyond the scope of this work. The power spectrum obtained numerically in this paper shows that within the parameter space we are concerned with, there is no noticeable effect from instabilities.




\section{Conclusions} \label{conclusion}

NEC violation could play a crucial role in the very early universe. Primordial GWs generated by inflation could carry rich information about the very early universe, making them an essential tool for exploring the physics related to NEC violation.
In this paper, we investigate the parity-violating effects in primordial GWs within two scenarios of the very early universe that involve NEC violation. These two scenarios are the bounce-inflation scenario and the scenario where intermediate NEC violation occurs during inflation.
For each scenario, we investigate the parity-violating effects produced when the background field and the spectator field are coupled with the Nieh-Yan term, respectively.


Using numerical methods, we solve for the power spectra of primordial GWs in different polarization modes. We find that when the background field $\phi$ is coupled with the Nieh-Yan term, the vigorous motion of $\phi$ around the NEC violation can significantly enhance the parity-violating effect. Moreover, the scale where this effect is most pronounced corresponds closely to the scale of the maximum of the primordial GW power spectrum. In other words, in the case where the background field is coupled with the parity-violating term, NEC violation naturally amplifies the observability of the parity-violating effect at scales where primordial GWs are most likely to be observed.
This is expected because the effective propagation speed of the primordial GWs for different polarization modes is close to the speed of light in this situation.
The locations where $P_T$ and $\Delta \chi$ reach their maximum values depend on the times at which $H$ and $\dot{\phi}$ reach their maximum values, respectively, which are very close for NEC violation of short duration.

In contrast, in the case where the spectator $\psi$ is coupled with the Nieh-Yan term, while $\psi$ can also enhance the parity-violating effect to some extent, the amplitude of the power spectrum at the scale where this effect is most significant is much smaller compared to its maximum value. This is because achieving a significant parity-violating effect from spectators necessitates a substantial increase in the effective comoving particle horizon of perturbation modes (see Sec. \ref{sec:why-spectator}), which in turn leads to the scale of the maximum parity-violating effect being much larger than the scale corresponding to the maximum of the power spectrum.

Therefore, spectators are unlikely to generate parity-violating effects of sufficient interest for observational purposes in the primordial GW background.
In other words, in NEC-violating early universe cosmological models, significant observable parity-violating effects in primordial GWs are primarily attributed to contributions from the background field, i.e., the physics directly related to NEC violation. This conclusion holds for NEC-violating scenarios, at least in the absence of excessive fine-tuning.
This result highlights the potential of primordial GWs as crucial tools for exploring NEC-violating and parity-violating physics.

We have discussed the evolution of primordial GW modes at super-horizon scales during NEC violation and analyzed the resulting blue-tilted spectrum features. We have also analyzed the reasons why spectators struggle to produce significant parity-violating effects at PTA scales (see Sec. \ref{sec:why-spectator}).
Since PTAs are blind to the circular polarizations of primordial GWs \cite{Kato:2015bye, Belgacem:2020nda, Cruz:2024esk}, it is interesting to investigate the anisotropy of the parity-violating primordial GW background produced by our models, which will be left for future work.

Our focus has been on the tensor sector of primordial perturbations.
In the two scenarios discussed in this paper, the power spectrum of primordial scalar perturbations generated by the background field $\phi$ also exhibits interesting observable features, while neither the Nieh-Yan term nor the spectator field has a significant impact on the primordial scalar perturbations at the quadratic order. The power spectrum of primordial scalar perturbations generated by bounce-inflation is suppressed at large scales with multipoles $\ell < 10$, while it is nearly scale-invariant at scales with $\ell > 10$. Therefore, it could better fit the CMB observational data (see, e.g., \cite{Cai:2017pga}).
In a model where there is an intermediate NEC violation during inflation, the power spectrum of primordial scalar perturbation is nearly scale-invariant at CMB scales. It is enhanced at intermediate scales due to the NEC violation, reaching a peak, and then slightly decreases at smaller scales while returning to near scale-invariance, although the amplitude can be larger than that of the power spectrum at CMB scales.
Such a power spectrum not only satisfies the observational constraints from the CMB but also has the potential to generate primordial black holes and scalar-induced GWs at certain scales \cite{Cai:2023uhc}.
These interesting features in the primordial scalar perturbations and GWs provide potential signatures for probing NEC violation in the very early universe.

\acknowledgments

We thank Mian Zhu, Jing Liu, Gen Ye, Tao Zhu, Xian Gao, Jun Zhang and Zu-Cheng Chen for stimulating discussions. Y. C. and Z.-W. J. are supported in part by the National Natural Science Foundation of China (Grant No. 11905224), the China Postdoctoral Science Foundation (Grant No. 2021M692942), the Natural Science Foundation of Henan Province and Zhengzhou University (Grant Nos. 242300420231, JC23149007, 35220136).
F. W. is supported by the National Natural Science Foundation of China (Grant No. 12075213) and the Natural Science Foundation for Distinguished Young Scholars of Henan Province (Grant No. 242300421046).
Y.-S. Piao is supported by the National Natural Science Foundation of China (Grant No. 12075246), National Key Research and Development
Program of China (Grant No. 2021YFC2203004), and the Fundamental Research Funds for the
Central Universities.
We acknowledge the use of the computing
server {\it Arena317}@ZZU.


\bibliography{Ref}
\bibliographystyle{utphys}

\end{document}